\documentclass[12pt,oneside]{article}
\begin{document}
%\frontmatter

%\topmargin=-.35in
%\textheight=8.60in
%\oddsidemargin=0.0in
%\textwidth=6.3in

%\begin{titlepage}
\begin{center}
{\large\bf A Brief Summary of the Group-Variation Equations\\}
%\vspace*{.7in}
{Chris Austin (chrisaustin@ukonline.co.uk)\\
%\vspace*{.3in}
33 Collins Terrace, Maryport, Cumbria CA15 8DL, England\\
}
\end{center}
%\vspace*{0.8in}
\begin{center}
{\bf Abstract}
\end{center}
\noindent A brief summary is given of the Group-Variation Equations and 
the island diagram confinement mechanism, with an explanation of 
the prediction that the cylinder-topology minimal-area spanning surface 
term in the correlation function of two Wilson loops at large $N_c$, 
when it exists, must have a pre-exponential factor, which for large area 
$A$ of the minimal-area cylinder-topology spanning surface, decreases 
with increasing $A$ at least as fast as $1/\ln(\sigma A)$, where 
$\sigma$ is the area law parameter.  This prediction is expected to be 
testable in lattice calculations.

\section{Description of the Group-Variation Equations}

A Group-Variation Equation \cite{GVE} expresses the derivative with respect to the
coupling constant $g$, of a coefficient in the $1/N$ expansion \cite{Planar} of a VEV or
correlation function of Wilson loops, in terms of a sum of modified Feynman
diagrams, related to the ordinary Feynman diagrams contributing to that VEV
or correlation function.

The modifications to the Feynman diagrams are:
\begin{itemize}
\item[(1)] Each window in an ordinary planar diagram is weighted by the VEV of the
Wilson loop that forms its perimeter.  The gluon and FP propagators are
first resolved into sums over paths, so that the window weights become
weights in the sums over paths.
\item[(2)] A planar diagram can have ``islands'', which look like planar vacuum
bubbles, drawn within its windows.  The non-simply connected window that
surrounds one or more islands, is weighted by the correlation function of
the two or more Wilson loops that form its perimeter.  Again, the gluon and
FP propagators are first resolved into sums over paths, so that the window
weights become weights in the sums over paths.
\item[(3)] Each planar diagram is multiplied by a numerical coefficient, which is
the derivative, at $M=1$, of the ``chromatic polynomial'', $\mathbf{C}(M)$, of
the diagram.  The chromatic polynomial, $\mathbf{C}(M)$, is by definition
the number of distinct ways of colouring the windows of the diagram, with
$M$ different colours available, subject to the ``map-colouring'' rule, that
no two windows that share a common border, can be coloured the same colour.
\end{itemize}

For vast classes of diagrams, $\mathbf{C}(M)$ has two or more factors of
$(M-1)$, so the numerical coefficient vanishes, and the diagram makes no
contribution.  In particular, no contributing diagram can have more than one
island, and if a diagram has an island, then it can have no propagators that
do not form part of that island.

In the RHS of the Group-Variation Equation
for the leading term in the $1/N$ expansion of the VEV of a single Wilson
loop, only two classes of diagrams survive:
\begin{itemize}
	\item[(a)] Non-island diagrams.  These have no island, and if you rub out the
Wilson loop, what remains must have only one connected component.
\item[(b)] Island diagrams.  These have exactly one island, and no propagators that
do not form part of the island.  Thus the window weight, for the one
non-simply connected window, consists of the correlation function of the LHS
Wilson loop, and the Wilson loop that forms the outer boundary of the
island, after the gluon and FP propagators in the island have been resolved
into sums over paths.
\end{itemize}

The Group-Variation Equations are closed and complete, in the sense that the
ordinary Feynman diagram expansions, for the VEVs and correlation functions,
can be recovered by developing their solutions in powers of $g$, with
appropriate ``boundary conditions'' at $g=0$, due to the derivative with
respect to $g$ in the LHS.

The equations for the coefficients at a common
non-vanishing order, in the $1/N$ expansions of the VEVs and correlation
functions, close among themselves.  In particular, the equations for the
leading non-vanishing terms in the $1/N$ expansions for the VEV of one
Wilson loop, and the correlation functions of two or more Wilson loops,
close among themselves.

By means of the Renormalization Group \cite{RGE}, the equations may be transformed into
equations for the derivative of the VEV or correlation functions with
respect to $\ln L$, where $L$ is a common scaling parameter, as the sizes
and separations, of all the Wilson loops involved in the VEV or correlation
function, are uniformly re-scaled to a different size.  The equations may
then be integrated with respect to $L$, starting from ``boundary conditions''
at small $L$, as given by renormalization-group improved perturbation
theory, and continuing to arbitrarily large $L$, because the correct
large-distance behaviour, specifically the Wilson area law \cite{Wilson} for the VEV of
one Wilson loop, and massive glueball saturation of the correlation
functions, solves the equations self-consistently at large $L$.

The derivation of the Group-Variation Equations is outlined in Section 
\ref{Derivation}.

\section{The Island Diagram Confinement Mechanism}

The demonstration that the correct large-$L$ behaviour solves the
Group-Variation Equation for the VEV of a single Wilson loop, at large $L$,
consists of the following observations:
\begin{itemize}
	\item[(i)] the principal effect of the window weights is to generate an effective
mass of at least $1.3\sqrt{\sigma}$ for each gluon and FP propagator, where
$\sigma$ is the area-law parameter.
\item[(ii)] the effective mass suppresses configurations with long propagators.
Thus each non-island diagram gives a contribution, at large $L$,
proportional to the perimeter of the LHS Wilson loop.
\item[(iii)] the contribution of each island diagram, at large $L$, is dominated by
the contributions of islands of a fixed size, of order
$\frac{1}{\sqrt{\sigma}}$.
\item[(iv)] the correlation function of a Wilson loop of the fixed size
$\frac{1}{\sqrt{\sigma}}$, and a Wilson loop of much larger size, of order
$L$, is strongly peaked at configurations where the smaller Wilson loop lies
close to the minimal-area spanning surface of the larger Wilson loop.
\item[(v)] near the configurations mentioned in (iv), the correlation function of
the two Wilson loops approximately factorizes into a factor equal to the VEV
of the large Wilson loop, and a factor that depends on the shape and
orientation of the small loop, and its perpendicular distance from the
minimal-area spanning surface of the large loop, but not on the position of
the perpendicular projection of the small loop, onto the minimal-area
spanning surface of the large loop.
\item[(vi)] the factor in (v) that depends on the shape and orientation 
of the small loop, and the perpendicular distance of the small loop from
the minimal-area spanning surface of the large loop, is expected to be
approximately independent of the shape and size of the large loop.  This 
is exactly true for the term in the correlation function that has
the form of the lightest glueball, propagating by the shortest path,
between the separate minimal-area spanning surfaces of the two loops, but
is slightly violated by the cylinder-topology minimal-area spanning 
surface term in the correlation function, when the cylinder-topology 
term exists.  (This is why the cylinder-topology term has to have a 
pre-exponential factor.)
\item[(vii)] thus each island diagram gives a contribution, at large $L$,
approximately equal to a constant, times the area of the minimal-area
spanning surface of the left-hand side Wilson loop, times the VEV of the
left-hand side Wilson loop.
\item[(viii)] in consequence of (ii) and (vii), the large-$L$ behaviour, of the
left-hand side Wilson loop, is completely determined by the island diagrams.
\item[(ix)] since the derivative, with respect to $\ln L$, of $e^{-\sigma aL^2}$,
where $a$ is the area of the minimal-area spanning surface of the left-hand
side Wilson loop, when $L=1$, is $-2\sigma aL^2e^{-\sigma aL^2}$, the
left-hand side and the right-hand side have the same dependence on $a$ and $L$.
\item[(x)] comparing the constant factors in the LHS and the RHS, $\sigma$ is found to
be equal to $\left|\frac{\beta(g)}{g}\right|$, times a power series in $g^2$, that begins with
a term independent of $g$.
\item[(xi)] each coefficient in the power series, is equal to $\sigma$, times a
numerical coefficient.  Thus $\sigma$ cancels out, and an equation for the
critical value of $g^2$, $g^2(\sqrt{\sigma})$, at which $g^2$ stops
evolving, is obtained.  This in turn fixes $\sqrt{\sigma}/\Lambda_s$.
\item[(xii)] $g^2$ stops evolving, at $g^2(\sqrt{\sigma})$, because the
large-distance behaviour, of all physical quantities, is completely
determined by islands of the fixed size, $\frac{1}{\sqrt{\sigma}}$.
\end{itemize}

It is not obvious that the sign of the island diagram contributions comes
out correct.  The net sign, of the contributions of the leading island
diagrams, has to be opposite to what would be obtained, if a scalar
propagator, rather than gluon and FP propagators, was involved.  I have
verified that the required sign reversal occurs in the simplest non-trivial
calculation, namely the change in the contribution of an island diagram,
when the renormalization point is changed.  The calculation exactly
parallels, step by step, the calculation of the leading $\beta$-function
coefficient, in a gauge-covariant background field method, so in this
instance, the sign change is a direct consequence of the sign of the $\beta$-function.

If the sign comes out correct for the net contributions of the island 
diagrams, at each order in the explicit powers of $g^2$ in
the right-hand sides of the Group-Variation Equations, then the island
diagram mechanism works in exactly the same way at each finite order, including at 
leading order, as it does for the full sum of all the terms in the
right-hand sides of the Group-Variation Equations.  Evidence that these
sums converge is given in Section \ref{Convergence}.

\section{The Pre-Exponential Factor in the Cylinder-\\Topology Term}

If the correlation function of two Wilson loops contains a term $e^{-\sigma
A}$, where $A$ is the area of the cylinder-topology minimal-area spanning
surface of the two loops, if it exists, then step (vi) goes slightly wrong.
The exponent that determines the rate of decrease of the correlation
function, as the small loop moves away from the minimal-area spanning
surface of the large loop, has a factor $\ln(\sigma A_L)$ in the 
denominator, where $A_L$ is the area of the minimal-area spanning 
surface of the large loop,
which results in the contribution of an island diagram getting an extra,
unwanted, factor of $\ln(\sigma A_L)$.  This extra factor of $\ln(\sigma A_L)$
would result in the VEV of a single Wilson loop decreasing slightly too fast
as the area, $A_L$, of its minimal-area spanning surface increased, and in fact,
violating the Seiler bound \cite{Seiler}, \cite{Simon Yaffe}, so it must be cancelled.  The simplest solution
is to assume that the cylinder-topology term, in the correlation function of
two Wilson loops, gets a pre-exponential factor, that decreases at least as
fast as $1/\ln(\sigma A)$, as $A$ increases.

The term in the correlation
function, that has the form of the lightest glueball, propagating by the
shortest possible path, between the separate minimal-area spanning surfaces
of the two loops, always gives a contribution of the correct form.

\section{The Group-Variation Equations for the Correlation Functions}

The demonstration that the correct large-$L$ behaviour solves the
Group-Variation Equations for the correlation functions of two or more
Wilson loops, at large $L$, is similar to the case of the VEV of a single
Wilson loop, in that the dominant contributions, to the right-hand side, are
given by the island diagrams.  Instead of minimal-area spanning surfaces,
the calculation involves the minimal-length spanning tree of the
configuration of well-separated loops.  A self-consistency requirement is
found, namely that the mass of the lightest glueball must be strictly less
than twice the effective mass generated for the gluon lines by the window
weights.

If the pre-exponential factor in the cylinder-topology term, in the
correlation function of two Wilson loops, decreases strictly faster than
$1/\ln(\sigma A)$, then the cylinder-topology term makes no contribution to
the leading behaviour, at large $L$, of the right-hand side of the
Group-Variation Equation for the VEV of a single Wilson loop.  In this case
$\sigma$, and the square of the mass $m_{0^{++}}$ of the lightest glueball,
are given by series whose terms differ only by simple numerical
coefficients, so that if the contributions of all but the leading order
island diagrams are neglected, and the ratio is taken, then the zeroth-order
estimate $m_{0^{++}}/\sqrt{\sigma}=2.38$ is obtained, which is about 33\%
less than the best lattice value of $3.56$ \cite{Teper}.

\section{Implication of the Lattice Value of $m_{0^{++}}$, for the 
Critical Value of $\alpha_s$}

Combining the zeroth-order estimate of $m_{0^{++}}/\sqrt{\sigma}$, with the
self-consistency constraint, that $m_{0^{++}}$ must be strictly less than
twice the effective mass generated for the gluon lines by the window
weights, does not provide any new information about the effective mass,
which is estimated in reference \cite{GVE} to be at least $1.3\sqrt{\sigma}$.  If we
use, instead, the best lattice value of $m_{0^{++}}$, then we find that the
effective mass must be strictly greater than $1.78\sqrt{\sigma}$.  If we
estimate the reciprocal of the typical island size, or in other words, the
mass at which $g^2$ stops evolving, more closely as twice the effective
mass, (since stretching an island, in any direction, elongates at least two
propagators), then we find that the mass, at which $g^2$ stops evolving,
must be strictly greater than $3.56\sqrt{\sigma}$.  Since the experimental
value of $\sqrt{\sigma}$ is about 0.44 GeV \cite{Michael}, this means that the mass, at
which $g^2$ stops evolving, must be strictly greater than 1.57 GeV, which is
not much smaller than $m_\tau$ = 1748 MeV, which is the smallest mass for
which $\alpha_s$ is known experimentally \cite{beta in MS bar 1}.
Another way of looking at this, is that the reciprocal of the typical
island size, which is the mass at which $g^2$ stops evolving, must be
strictly greater than the mass of the lightest glueball.

\section{Convergence of the Sums in the Right-Hand \\Sides of the
Group-Variation Equations}
\label{Convergence}

't Hooft has demonstrated, in reference \cite{'t Hooft a}, that the sums 
of the planar Feynman diagrams, in
large-$N_c$ QCD, converge geometrically, for a sufficiently small value of
$g^2$, if one throws away all the divergent subdiagrams, and furthermore,
in reference \cite{'t Hooft b}, that a similar result holds in the 
presence of the divergent subdiagrams, if
one uses a suitably generalized running coupling, and gives the gluons a
mass, to cut off the large-distance growth of the running
coupling.  It is therefore reasonable to suppose that the sums in the
right-hand sides of the Group-Variation Equations will converge
geometrically, for a sufficiently small value of $g^2$.  It would also seem
reasonable to suppose that in a natural renormalization scheme, such as $\overline{MS}$ \cite{MS bar}, the convergence behaviour of the large-$N_c$ limit of $\frac{\beta(g)}{g}$, as
a power series in $g^2$, will be neither better, nor worse, than the
convergence behaviour of the sums in the right-hand sides of the
Group-Variation Equations.  The large-$N_c$ limit of $\frac{\beta(g)}{g}$, as a
power series in $g^2$, may therefore also be expected to converge
geometrically, for a sufficiently small value of $g^2$.  The known expansion
coefficients, in the large-$N_c$ limit of $\frac{\beta(g)}{g}$, all have the same
sign, in $\overline{MS}$, and it is reasonable to expect this trend to continue.  We
may therefore expect that the direction of fastest growth, in the complex
$g^2$ plane, of the large-$N_c$ limit of 
$\left|\frac{\beta(g)}{g}\right|$, in $\overline{MS}$, will be
along the positive real axis.  The critical value of $g^2$ is essentially
determined, by (x) and (xi) above, as the point at which 
$\left|\frac{\beta(g)}{g}\right|$
reaches a critical value.  Therefore if these suppositions are correct, 
the critical value of $g^2$ is 
strictly smaller than the radius of convergence of the large-$N_c$ limit 
of $\frac{\beta(g)}{g}$, in $\overline{MS}$, as a power series in $g^2$, 
so the large-$N_c$ limit of $\frac{\beta(g)}{g}$, in $\overline{MS}$, 
converges geometrically, as a power series in $g^2$, at the critical 
value of $g^2$, and the sums in the right-hand sides of the 
Group-Variation Equations, in $\overline{MS}$, also converge
geometrically, at the critical value of $g^2$.

\section{Implication of the $\beta$-Function to Four Loops}

Study of the large-$N_c$ limit of the general result for $\frac{\beta(g)}{g}$, in
$\overline{MS}$, to four loops, given in reference \cite{beta in MS bar 
2}, shows that the ratios of successive pairs of
coefficients in the expansion are increasing, but at a decreasing rate, and
indicates that the series is likely to diverge for a value of $g^2$ that
corresponds to $\alpha_s$ lying somewhere in the range 0.43 to 0.85, and
most likely, near the lower end of this range.  Since the critical value of
$\alpha_s$ will be strictly less than the value of $\alpha_s$, that
corresponds to the value of $g^2$ at which the series diverges, and
$\alpha_s(m_\tau)$ is equal to 0.35 \cite{beta in MS bar 1}, this is further evidence that the
critical value of $\alpha_s$ is not much larger than $\alpha_s(m_\tau)$.

\section{Derivation Of The Group-Variation Equations}
\label{Derivation}

The Group-Variation Equations are derived by expressing the VEVs and
correlation functions of $\textrm{SU}(NM)$ Yang-Mills theory, first in terms of the
VEVs and correlation functions of $(\textrm{SU}(N))^M$ Yang-Mills theory, using
general expansions which express the VEVs and correlation functions of a
group, in terms of those of a subgroup, then in terms of the VEVs and
correlation functions of $\textrm{SU}(N)$ Yang-Mills theory, using the factorization
properties of the VEVs and correlation functions of $(\textrm{SU}(N))^M$ Yang-Mills
theory.  Substituting in the $1/N$ expansions of the VEVs and correlation
functions, and equating coefficients of powers of $1/N$, equations for the
coefficients in the $1/N$ expansions of the VEVs and correlation functions
are obtained, in which the $M$-dependence of the left-hand sides is through
an overall power of $M$, and the replacement of the coupling constant $g^2$ by $g^2M$,
and the $M$-dependence of the right-hand sides is through the chromatic
polynomial factors $\mathbf{C}(M)$ of the diagrams.  Taking the derivative
with respect to $M$, at $M=1$, and in the left-hand sides, expressing the
derivative with respect to $M$, in terms of the derivative with respect to
$g^2$, gives the Group-Variation Equations.

\vspace{0.5cm}

I would like to thank D. Leinweber, J. Negele, and H. Shanahan for 
emails about the possibility that the pre-exponential factor, in the 
cylinder-topology term, can be tested in lattice calculations, B. Alles 
for an email noting the practical difficulties of such a test, and M. 
Karliner and Z. Sroczynski for suggesting I write a brief summary.

\end{document}